\documentclass[superscriptaddress,twocolumn,showpacs,preprintnumbers,prb]{revtex4}
\usepackage{graphicx}
\usepackage{times}
\usepackage{epsfig}
\begin{document}

\def\salto{\vskip 1cm}
\def\lgr{\langle\langle}
\def\rgr{\rangle\rangle}

\newcommand{\UT} {Condensed Matter Sciences Division, Oak Ridge National Laboratory, Oak Ridge, 
TN 37831 and Department of Physics, University of Tennessee, Knoxville, TN 37996}
\newcommand{\FSU}{National High Magnetic Field Laboratory and Department of Physics, Florida 
State University, Tallahassee, FL 32306}
\newcommand{\OU} {Department of Physics, Oakland University, Rochester, MI 48309}

\title{Electron Transport through a Molecular Conductor with Center-of-Mass Motion }

\author{K.A. Al-Hassanieh }\affiliation {\UT}\affiliation {\FSU} 
\author{C.A. B\"usser }    \affiliation {\UT}
\author{G.B. Martins }     \affiliation {\OU}
\author{E. Dagotto }       \affiliation {\UT}

\begin{abstract}
The linear conductance of a molecular conductor oscillating between two metallic leads is investigated 
numerically both for Hubbard interacting and noninteracting electrons.  The molecule-leads tunneling 
barriers depend on the molecule displacement from its equilibrium position.  The results present an 
interesting interference which leads to a conductance dip at the electron-hole symmetry point, that 
could be experimentally observable.  It is shown that this dip is caused by the destructive interference 
between the purely electronic and phonon-assisted tunneling channels, which are found to carry opposite 
phases.  When an internal vibrational mode is also active, the electron-hole symmetry is broken but a 
Fano-like interference is still observed.
\end{abstract}

\maketitle
Molecular electronics has received much attention in the past decade, particularly since it became 
possible to fabricate devices in which the active element is a single molecule.\cite {Ratner,Reed}  
A fundamental property of molecular conductors is their discrete electronic spectrum.  Although the weak 
coupling of the molecule to the two metallic electrodes leads to the broadening of the molecular energy 
levels, their discrete nature is maintained.  Due to the small size of these molecules, electronic 
correlations are dominant and they lead to interesting many-body effects, such as the Coulomb blockade 
and Kondo resonance\cite{Glazman}.  These effects have been observed experimentally in molecular 
conductors\cite{JPark} and other nanostructures.\cite {Goldhaber}  Another interesting property of 
molecules is their flexible nature.  They have an intrinsic spectrum of internal vibrational modes and, 
when coupled to the electrodes, some molecules acquire external vibrational modes as well.  The excitation 
of one or more of these modes leads to the modulation of the electronic energy levels
and tunneling barriers between the molecule and the electrodes or between different parts of the molecule, 
thus changing the molecular transport properties.  These vibrational effects have been observed in a number 
of recent experiments,\cite{HPark,Natelson} and have been the subject of considerable theoretical 
investigation\cite{Mitra}.

In this work, we study the zero bias conductance of a molecular conductor model with one relevant electronic 
energy level, both with interacting and noninteracting electrons.  The molecule is allowed to oscillate 
between the two electrodes.  This center-of-mass (CM) vibrational mode is treated quantum mechanically and 
leads to an asymmetric modulation of the tunneling barriers molecule-electrodes.  The vibrational excitation 
is also coupled to the charge on the molecule.  This is due to the fact that the chemical bonds inside the 
molecule and the molecule-electrode bonds depend in general on the molecule's charge.  The results show an 
interesting and unexpected conductance cancellation when an odd number of electrons occupy the molecule.  
It is discussed below that this cancellation is due to the {\it destructive interference between the purely 
electronic and phonon-assisted tunneling channels}, which are found to carry opposite phases.
\cite{Silva,Khaled}  In this case both channels are elastic.  The phonons are virtual, not thermal. 
\begin{figure}
\vspace{0.cm}
\epsfxsize=6cm \centerline{\epsfbox{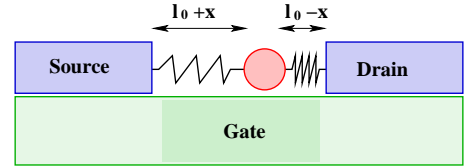}}
\caption{A schematic of the system studied in this paper.  The molecule can oscillate between the two leads around 
the equilibrium position $\rm l_0$, thus modulating the molecule-leads tunneling barriers.}
\label{system}
\vspace{0.cm}
\end{figure}             

Figure \ref{system} schematically depicts the system analyzed in this work.  The molecule can oscillate 
between the source and drain electrodes, thus modulating the tunneling barriers.  In our calculations, this 
modulation and the electron-vibration coupling are expanded up to the linear term.\cite{note,Liliana}  The 
electronic part of the system is modelled using the Anderson impurity Hamiltonian.  The total Hamiltonian 
can be written as $\hat H = \hat H_{\rm M} + \hat H_{\rm {leads}} + \hat H_{\rm{M-leads}}$, where 
$\hat H_{\rm M}$ is the Hamiltonian of the molecule, 
\begin{equation}
\hat H_{\rm M} = V_g \hat n_d + U\hat n_{d \uparrow}\hat n_{d \downarrow} + \lambda (1-\hat n_d)(a+a^\dagger) 
+ \omega_0 a^\dagger a.
\end{equation}
The first term represents the energy of the relevant molecular orbital controlled by the gate voltage,  
the second term represents the Coulomb repulsion between the electrons occupying the molecular orbital, 
the third term couples the vibrational excitation to the net charge on the molecule ($a^\dagger$ and $a$ are 
the phonon creation and annihilation operators), and the fourth term represents the vibrational energy.  
$\hat H_{\rm{leads}}$ describes the two leads modeled here as semi-infinite ideal chains,      
\begin{equation}
\hat H_{\rm{leads}} = -t\sum_{i\sigma} (c_{li\sigma}^{\dagger} c_{l i+1\sigma} + c_{ri\sigma}^{\dagger} 
c_{r i+1\sigma} + h.c.),
\end{equation}
where $c_{li\sigma}^{\dagger}$ $(c_{ri\sigma}^{\dagger})$  creates an electron with spin $\sigma$ at site $i$ 
in the left (right) lead.  $t$ is the hopping amplitude in the leads and the energy scale $(t = 1)$.
$\hat H_{\rm{M-leads}}$ connects the molecule to the leads, 
\begin{eqnarray}
\hat H _ {\rm{M-leads}} &=& t'[1 - \alpha (a+a^\dagger)]\sum_{\sigma} ~(d_{\sigma}^\dagger c_{l0\sigma} + h.c.) 
 ~+\nonumber \\
&& t'[1 + \alpha (a+a^\dagger)] \sum_{\sigma}~(d_{\sigma}^\dagger c_{r0\sigma} + h.c.),
\label{shuttle1}
\end{eqnarray}
where $d_{\sigma}^{\dagger}$ creates an electron with spin $\sigma$ in the molecule, $t'$ is the hopping 
parameter between the molecule and the first site of each lead, and $\alpha$ is a parameter that carries the 
dependence of $t'$ on the molecule displacement from its equilibrium position $\hat x$ (note the opposite signs 
in this dependence for the two leads).  This displacement can be written in terms of the phonon operators as 
$\hat x = (a + a^\dagger)$.  The total Hamiltonian is invariant under the particle-hole and ($a \to -a$) 
transformation. In the results shown, unless otherwise stated the following set of parameters was used 
$(U = 1.0, t' = 0.2, \omega_0 = 0.2)$ while $\lambda$ and $\alpha$ were varied.\cite {Busser}The value of $\omega_0$ 
was fixed since its increase or decrease would simply produce the opposite effect of increasing or decreasing $\lambda$ 
and/or $\alpha$. 
\begin{figure}
\vspace{0.cm}
\epsfxsize=8cm \centerline{\epsfbox{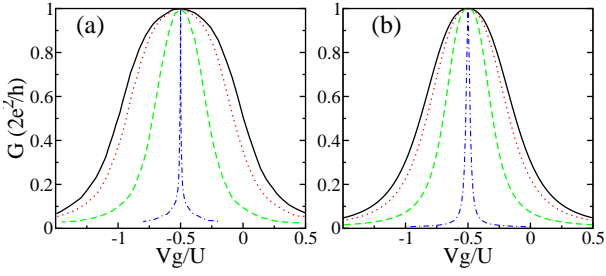}}
\caption{(a) NRG and (b) ED+DE results for G as a function of $V_g$ in the Kondo Regime for $\alpha = 0$ and increasing 
the electron-phonon coupling strength ($\lambda/\omega_0 = 0.0, 0.4, 0.8,$ and $1.2$).  For small $\lambda/\omega_0$, 
the standard Kondo resonance is obtained with renormalized parameters.  In the limit of large $\lambda/\omega_0$, the 
charge Kondo effect is obtained.}
\label{results0}
\vspace{0.cm}
\end{figure}  

Using the Keldysh formalism,\cite{Meir} the zero bias and zero temperature conductance can be written as 
$ G = {2e^2\over h}|t^2G_{lr}(\epsilon_F)|^2[\rho(\epsilon_F)]^2$, where $G_{lr}$ is the Green's
 function that propagates an electron from the left to the right lead and $\rho(\epsilon_F)$ is the density of 
states in the leads at the Fermi level. Note that at zero bias only elastic processes can be observed.  The 
Green's functions are calculated using exact diagonalization supplemented by a Dyson equation embedding procedure 
(ED+DE).  (See Ref.[\onlinecite{method}] for a full description of the method).

It is useful to start by studying briefly the model when the tunnel barriers do not depend on the vibrational 
excitation ($\alpha=0$).  This case was previously studied using numerical renormalization group (NRG) techniques.\cite{Cornaglia1}  
In this case, the model can be mapped into an effective electronic Hamiltonian with renormalized parameters ($\tilde U=U-
2\lambda^2/\omega_0$, $\tilde V_g = V_g+\lambda^2/\omega_0$ and $\tilde t'\propto t'exp({-\lambda^2\over 2\omega_0^2})$)
\cite{Hewson,Cornaglia1}.  Figures \ref{results0}a and b show the results in the interacting electrons case obtained using 
ED+DE and NRG, respectively.  For weak electron-phonon coupling, the standard Kondo resonance 
is obtained with renormalized parameters.  In the strong electron-phonon coupling limit, the 'charge Kondo effect' is 
obtained.  Note that the results obtained using ED+DE are very similar to the NRG results, clearly capturing the essence 
of the problem.  Since NRG methods cannot be applied to the CM oscillations studied here, and ED+DE appears equally 
accurate, the results presented below were obtained using this last technique.

\begin{figure}
\vspace{0.cm}
\epsfxsize=7cm \centerline{\epsfbox{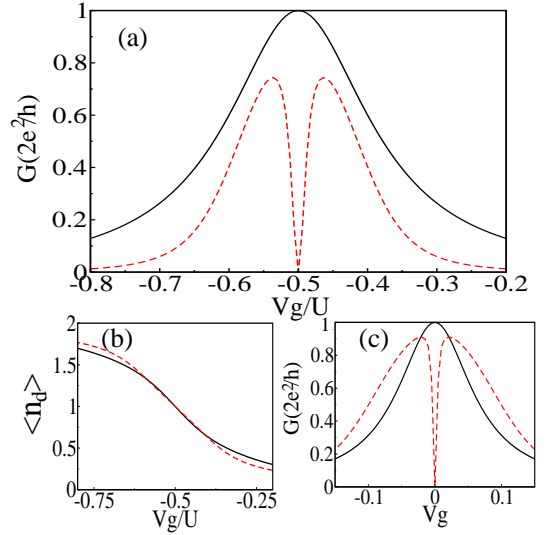}}
\caption{ (a) G as a function of $V_g $ in the Kondo regime for $\alpha = 0$ (solid line) and $\alpha = 0.4$ (dashed 
line), $\lambda  = 0.2$ in both cases.  In the first case, the usual Kondo peak with reduced width is obtained.  In the second case, 
a conductance dip is obtained.  
(b) Average occupation of the molecular level for the same set of parameters. Note that
the charging behavior is almost the same in both cases.  (c)  G vs. $V_g$ in the absence of Coulomb repulsion ($U=0$),
for $\alpha = 0$ (solid line) and $\alpha = 0.4$ (dashed line), $\lambda = 0.1$ in both cases.  The conductance dip is also 
obtained in this case, thus the physical mechanism behind this effect does not depend on the electron-electron interaction.}
\label{results1}
\vspace{0.cm}
\end{figure}
Figure \ref{results1} contains the main results of this work.  Figure \ref{results1}a shows the conductance in the interacting 
 electrons case.  For $\alpha = 0$, the conductance simply shows a Kondo resonance
peak with reduced width.  However, when $\alpha \neq 0$, a conductance {\it dip} is obtained when an odd number of 
electrons occupy the molecule.  Figure \ref{results1}b provides the average occupation $\langle n_d \rangle $ of the 
molecular orbital where it can be clearly seen that the charging behavior is almost the same in both cases.  Note that 
for $\alpha \neq 0$, the usual Friedel sum rule \cite{Hewsonbook} $G = {{2e^2}\over h}\sin^2 ({\pi\over 2}\langle n_d \rangle)$ 
is not satisfied and this can be an indication of a non-Fermi liquid behavior.  Figure \ref{results1}c shows the conductance 
in the absence of Coulomb repulsion $(U = 0)$: the same effect is obtained as for the case $\alpha \neq 0$.  The conductance 
cancellation does not depend on the electron-electron interaction, which agrees with the qualitative explanation presented 
below.

Figure \ref{results2} shows the conductance as a function of $V_g$ for different values of $\lambda$.  The dip becoming more pronounced 
as $\lambda$ increases i.e. as the average number of phonons in the ground state increases.  
The inset shows the results obtained by truncating the phonon Hilbert space at different maximum number of phonons $(N_{\rm ph})$.  
In all the calculations, $N_{\rm ph} = 7$ was used unless stated otherwise.  The qualitative effect of conductance cancellation is 
obtained all the way down to $N_{\rm ph} = 1$, allowing us to study larger clusters and reduce size effects to intuitively 
understand the origin of the dip.   
\begin{figure}
\vspace{0.cm}
\epsfxsize=7.0cm \centerline{\epsfbox{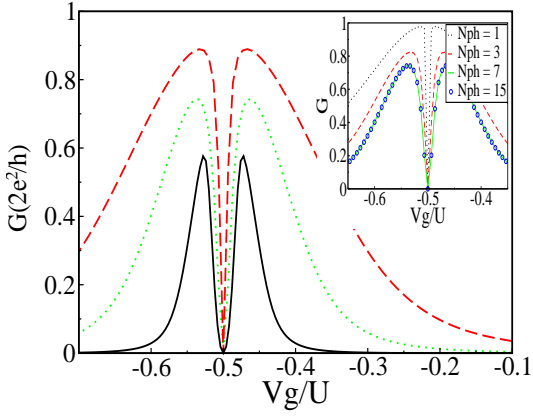}}
\caption{ G as a function of $V_g$ in the Kondo regime for $\alpha = 0.4$ 
and $\lambda = $0.15, 0.20, and 0.25 (dashed, dotted and solid lines respectively).  The dip becomes more pronounced as $\lambda$ 
increases.  The inset shows the convergence of the results with the maximum number of phonons ($N_{\rm ph}$).  Note that the 
qualitative effect of conductance cancellation is preserved all the way down to $N_{\rm ph} = 1$.}
\label{results2}
\vspace{0.cm}
\end{figure}

\begin{figure}
\vspace{0.cm}
\epsfxsize=7.5cm \centerline{\epsfbox{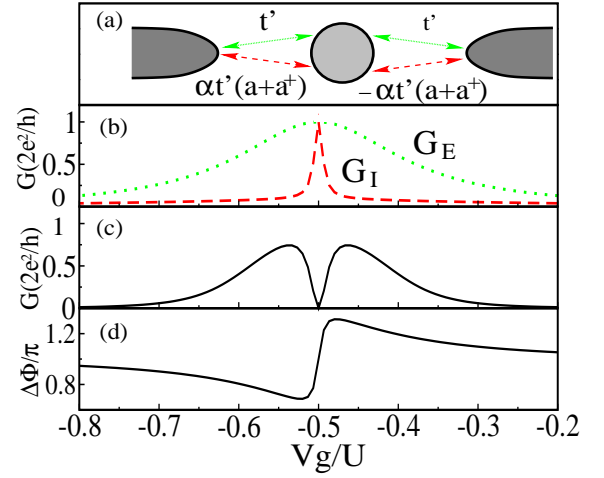}}
\caption{(a) A schematic representation of the two conductance channels, the purely electronic tunneling represented by the two 
upper arrows  and the ``phonon-assisted tunneling'' channel represented by the two lower arrows.  (b) Partial conductance when only 
one of the channels is active.  The dotted line shows the conductance of the purely electronic tunneling channel $G_{\rm E}$, 
while the dashed line shows the conductance of the ``phonon-assisted tunneling'' channel $G_{\rm I}$.  (c)  Conductance when both 
channels are active.  (d) Phase difference $\Delta\Phi$ of the two channels.  Note that $\Delta\Phi$ for all
values of $V_g$ is close to $\pi$ thus leading to destructive interference.  In particular, for $V_g = -U/2$, $\Delta\Phi = \pi$
and $G_{\rm E} = G_{\rm I} = 2e^2/h$, thus leading to a perfect cancellation in the overall conductance.  ($\lambda = 0.2$ , 
$\alpha = 0.4$).}
\label{explanation}
\vspace{0.cm}
\end{figure}

In figure \ref{explanation}, an explanation of the conductance dip is presented.  The reasoning starts by noting  
that $~\hat H_{\rm M-leads}~$ in Eq.\ref{shuttle1} can be rewritten as a sum of two channels  contributing to the overall 
molecule-leads connection.  The first term, $t'\sum_{\sigma} ~(d_{\sigma}^\dagger c_{l0\sigma} +d_{\sigma}^\dagger c_{r0\sigma} 
+ h.c.)$, represents the purely electronic tunneling between the molecule and the two electrodes. The second term, $~t'\alpha(a + 
a^\dagger)\sum_{\sigma} ~(d_{\sigma}^\dagger c_{r0\sigma} - d_{\sigma}^\dagger  c_{l0\sigma} + h.c.) $~, represents a phonon assisted 
tunneling channel, i.e. the electron absorbs (emits) a phonon upon entering the molecule and, then, emits (absorbs) a phonon upon 
leaving.  Note that both channels are coherent and elastic.  The number of phonons in the system does not change.  Fig. 
\ref{explanation}a shows a schematic of the two channels.  These channels were studied separately by keeping only the relevant term 
in $\hat H_{\rm M-leads}$.  The conductance and the phase carried by each channel were calculated.  Fig. \ref{explanation}b shows the 
conductance of the separate channels.  Fig. \ref{explanation}c contains the conductance when both channels are active i.e. when both 
terms are included in $\hat H_{\rm M-leads}$.  Fig. \ref{explanation}d shows the phase difference between the two channels.  Note that 
for $V_g = -U/2$, the conductance of each of the channels is $2e^2/h$ and the phase difference is $\pi$, leading to a perfect 
cancellation in the overall conductance.\cite {Khaled}  This interference effect is independent of the electron-electron interaction 
and, thus, the cancellation should still be present for $U = 0$ as already shown.  The dip becomes more pronounced as $\lambda$ 
increases, increasing the average number of phonons in the ground state.

\begin{figure}
\vspace{0.cm}
\epsfxsize=6cm \centerline{\epsfbox{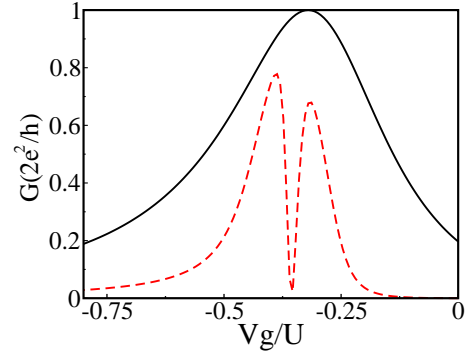}}
\caption{The solid line shows the conductance of the molecule when a breathing vibrational mode is active (no CM motion).  
The particle-hole symmetry is broken as expected and no conductance dip is obtained.  The dashed
line shows the conductance when both breathing and CM vibrational modes are active.  The combined effects of the two modes lead 
to a Fano-like interference.  The breathing mode parameters used are $\lambda' = 0.2$, $\omega'_0 = 0.3$ and $\alpha' = 0.3$.}
\label{breath}
\vspace{0.cm}
\end{figure}

The stability of the dip is tested by adding an internal vibrational mode which leads to the symmetric modulation of the tunnel 
barriers to the leads (breathing mode).  To account for this mode, the following terms were added to the Hamiltonian 
\begin{eqnarray}
 \hat H' &=& \lambda'(1 - \hat n_d)(b + b^\dagger) + \omega'_0 b^\dagger b 
 ~+\nonumber \\
&& t'\alpha' (b + b^\dagger)\sum_{\sigma} ~(d_{\sigma}^\dagger c_{l0\sigma} + d_{\sigma}^\dagger c_{r0\sigma} + h.c.), 
\label{secondmode}
\end{eqnarray}
where the first term represents the electron-phonon coupling, the second term represents the breathing vibrational energy, 
and the third term represents the subsequent modulation of the tunnel barriers.  The results are shown in Fig.\ref{breath}.  
When only the internal mode is active (solid line), the electron-hole symmetry is 
broken but no dip is observed.  This agrees with previous results\cite{Cornaglia2} obtained using NRG calculations.  In the case 
where both vibrational modes are active (dashed line), the dip appears.  The combined effect of conductance cancellation and 
electron-hole asymmetry leads to a Fano-like interference.    

\begin{figure}
\vspace{0.5cm}
\epsfxsize=8.0cm \centerline{\epsfbox{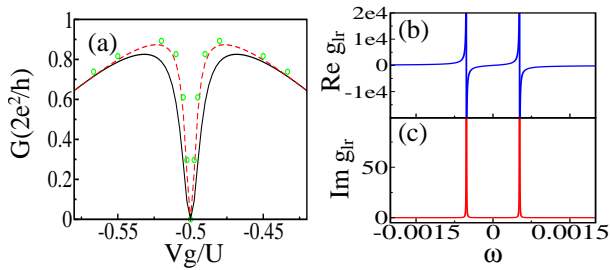}}
\caption{ (a) Convergence of the conductance with the size of the exactly-solved cluster $L$.  The solid line, dashed line, and 
the circles show the results obtained using $L = 3, 7,$ and $11$ respectively.  In the three cases, a maximum number of three 
phonons was used.  (b) The real and (c) imaginary parts of the isolated cluster Green's function $g_{\rm lr}$ that propagates 
an electron from the left to the right ends of the cluster ($L = 3$ and $V_g = -U/2$).  Note that both parts are equal to zero 
at the Fermi level (located at $\omega = 0.0$).  Thus, the origin of the conductance dip can be traced back to the exactly-solved 
cluster.  $\lambda = 0.2$ and $\alpha = 0.4$ in the three figures.}
\label{cluster}
\vspace{0.0cm}
\end{figure} 
The finite-size effects on the results are shown in Fig. \ref{cluster}a where the convergence of the conductance with
the size of the exactly-solved cluster is presented.  Note that increasing the cluster does {\it not} change the qualitative 
effect of the conductance dip. Moreover, the origin of the conductance dip can be traced back to the exactly-solved cluster 
by studying Green's functions {\it before} the embedding process.  Figures \ref{cluster}b and c show the real and imaginary 
parts of the Green's function $g_{\rm lr}$ that propagates an electron from the left to the right end of the cluster for 
$\alpha \neq 0$ and $V_g = -U/2$.  Both parts are zero at the Fermi level $(\omega = 0)$.  For $\alpha = 0$ (not shown here), 
$g_{\rm lr}$ has a pole at the Fermi level and the system is perfectly conducting.  When $\alpha$ is turned on, the pole splits 
into two, one below and one above the Fermi level thus causing the zero conductance.   

In conclusion, the zero temperature electron transport through a molecular conductor with center-of-mass motion was studied 
numerically for interacting and noninteracting electrons.  The results present an interesting conductance dip when an odd number 
of electrons occupy the molecule.  It is argued that this dip is caused by the destructive interference between the purely electronic 
and phonon-assisted tunneling channels, which were found to carry opposite phases.  When an internal vibrational mode is also active, 
the particle-hole symmetry is broken but a Fano-like interference is still obtained. The conductance cancellation would best be 
observed on a broad conductance peak such as the Kondo peak which is broader than the resonant tunneling peak.      

The authors thank E.V. Anda, S. Ulloa, P.S. Cornaglia and D. Natelson for discussions.  This effort has been partially supported by 
the NSF grant DMR-0454504.


\begin{thebibliography}{99}

\bibitem{Ratner} J.R. Heath and  M.A. Ratner, Physics Today, {\bf 43} May 2003.

\bibitem{Reed}
C. Joachim {\it et al.}, Phys. Rev. Lett. {\bf 74}, 2102 (1995); M.A. Reed {\it et al.}, Science {\bf 278}, 252 (1997); 
R.H.M. Smit {\it et al.}, 
Nature {\bf 419}, 906 (2001); J. Reicher {\it et al.}, Phys. Rev. Lett. {\bf 88}, 176804 (2002).    

\bibitem{Glazman} L.I. Glazman and M.E. Raikh, Pis'ma Zh \'Eksp. Teor. Fiz.{\bf 47}, 378 (1988) [JETP Lett. {\bf 47}, 452 (1988)];
T.K. Ng and P.A. Lee, Phys. Rev. Lett. {\bf 61}, 1768 (1988); Y. Meir and N.S. Wingreen, Phys. Rev. Lett. {\bf 68}, 2512 (1992);  
J. Paaske {\it et al.}, Phys. Rev. B {\bf 70}, 155301 (2004). 

\bibitem{JPark} J. Park {\it et al.},  Nature {\bf 417}, 722 (2002); W. Liang {\it et al.}, Nature {\bf 417}, 725 (2002); 
Lam H. Yu {\it et al.}, 
Nano Letters {\bf 4}, 79 (2004).

\bibitem{Goldhaber} D. Goldhaber-Gordon {\it et al.}, Nature {\bf 391}, 156 (1998); S.M. Cronenwett {\it et al.}, Science {\bf 281}, 
540 (1998);
V. Madhavan {\it et al.}, Science {\bf 280}, 567 (1998).  J. G\"ores {\it et al.}, Phys. Rev. B {\bf 62}, 2188 (2000).
 
\bibitem{HPark} H. Park {\it et al.}, Nature {\bf 407}, (2000) 57. 

\bibitem{Natelson} L.H. Yu {\it et al.}, Phys. Rev. Lett. {\bf 93}, 266802 (2004). 

\bibitem{Mitra} A. Mitra {\it et al.}, Phys. Rev. B {\bf 69}, 245302 (2004), and references therein; E. Vernek {\it et al.}, Phys. Rev. B 
{\bf 72}, 121405(R) (2005).

\bibitem{Silva} A. Silva {\it et al.}, Phys. Rev. B {\bf 66}, 195316 (2002); J.L. D'Amato {\it et al.} Phys. Rev. B {\bf 39}, 
3554 (1989); T.-S. Kim and S. Hershfield, Phys. Rev. B. {\bf 67}, 235330 (2003); W. Hofstetter and H. Schoeller, Phys. Rev. Lett. 
{\bf 88}, 016803 (2002); T.-S. Kim and S. Hershfield, Phys. Rev. B {\bf 63}, 245326 (2001).
                                                                                
\bibitem{Khaled} C.A. B\"usser {\it et al.}, Phys. Rev. B {\bf 70}, 245303 (2004). 

\bibitem{note} Some results were gathered keeping up to the second order in the tunneling barriers dependence on the molecule 
position, and the conclusions are unchanged.

\bibitem{Liliana} L. Arrachea {\it et al.} Phys. Rev. B {\bf 67}, 134307 (2003).

\bibitem{Busser} The value of $U$ was selected to allow for the Kondo cloud to fit in our finite lattice (see C.A. B\"usser {\it et al.}, 
Phys. Rev. B {\bf 62}, 9907 (2000)).  

\bibitem{Meir} Yigal Meir {\it et al.}, Phys. Rev. Lett. {\bf 66}, 3048 (1991); E.V. Anda and F. Flores, J. Phys.:Condens. 
Matter {\bf 3}, 9087 (1991); H.M. Pastawski, Phys. Rev. B {\bf 46}, 4053 (1992).  

\bibitem{method} A complete description of the method is available online (http://correlate8.phys.utk.edu/transport.pdf).  This 
method was originally proposed by E.V. Anda and G. Chiappe.  See V. Ferrari {\it et al.}, Phys. Rev. Lett.{\bf 82}, 5088 (1999); 
M.A. Davidovich {\it et al.}, Phys. Rev. B {\bf 65}, 233310 (2002); M.E. Torio {\it et al.}, Phys. Rev. B {\bf 65}, 
085302 (2002); A.A. Aligia {\it et al.}, J. Phys.:Condens. Matter {\bf 17}, S1095 (2005). 

\bibitem{Cornaglia1} P.S. Cornaglia {\it et al.}, Phys. Rev. Lett. {\bf 93}, 147201 (2004).

\bibitem{Hewson} A.C. Hewson and D. Meyer, J .Phys.:Condens. Matter {\bf 14}, 427 (2002).

\bibitem{Hewsonbook} A.C. Hewson, {\it The Kondo Problem to Heavy Fermions} (Cambridge University Press, 1997).
\bibitem{Cornaglia2} P.S. Cornaglia {\it et al.}, Phys. Rev B {\bf 71}, 075320 (2005).

\end{thebibliography}
\end{document}